\documentclass{article}
\usepackage[latin1]{inputenc} 
\usepackage{amssymb}
\usepackage[usenames]{color}

\newcommand{\col}[1]{\textcolor{red}{#1}}
\newcommand{\colv}[1]{\textcolor{green}{#1}}

\newcommand{\q}{\mathsf{q}}

\newcommand{\ql}{\mathsf{li}}
\newcommand{\qu}{\mathsf{un}}

\newcommand{\gl}{\mathsf{gl}}
\newcommand{\lo}{\mathsf{lo}}

\newcommand{\La}{\langle}
\newcommand{\Ra}{\rangle}
\newcommand{\Lc}{[}
\newcommand{\Rc}{]}
\newcommand{\ep}{\langle\rangle}
\newcommand{\Let}{\mathsf{let}}
\newcommand{\In}{\mathsf{in}}
\newcommand{\Letrec}{\mathsf{letrec}}
\newcommand{\If}{\mathsf{if}}
\newcommand{\Then}{\mathsf{then}}
\newcommand{\Else}{\mathsf{else}}

\newcommand{\True}{\mathsf{true}}
\newcommand{\False}{\mathsf{false}}

\newcommand{\Bool}{\mathsf{bool}}
\newcommand{\Int}{\mathsf{int}}

\newcommand{\New}{\mathsf{new}}
\newcommand{\At}{\mathsf{at}}

\newcommand{\T}{\mathsf{T}}

\newcommand{\va}{\mathsf{v}}
\newcommand{\op}{\mathsf{o}}
\newcommand{\e}{\mathsf{e}}

\newcommand{\p}{\mathsf{p}}
\newcommand{\pp}{\overline{\mathsf{p}}}

\newcommand{\x}{\mathsf{x}}

\newcommand{\y}{\mathsf{y}}

\newcommand{\z}{\mathsf{z}}
\newcommand{\zl}{\mathsf{\zeta}}
\newcommand{\w}{\mathsf{w}}

\newcommand{\n}{\mathsf{n}}
\newcommand{\f}{\mathsf{f}}
\newcommand{\h}{\mathsf{h}}
\newcommand{\g}{\mathsf{g}}
\newcommand{\vi}{\mathsf{i}}

\newcommand{\vj}{\mathsf{j}}
\newcommand{\vr}{\mathsf{r}}

\newcommand{\vk}{\mathsf{k}}

\newcommand{\vb}{\mathsf{b}}
\newcommand{\vd}{\mathsf{d}}
\newcommand{\bl}{\mathsf{\beta}}
\newcommand{\dl}{\mathsf{\delta}}

\newcommand{\m}{\mathsf{m}}

\newcommand{\xs}{\mathsf{xs}}
\newcommand{\D}{\mathsf{D}}
\newcommand{\V}{\mathsf{V}}
\newcommand{\B}{\mathsf{B}}
\newcommand{\C}{\mathsf{C}}
\newcommand{\A}{\mathsf{A}}
\newcommand{\Al}{\mathsf{\alpha}}\newcommand{\R}{\mathsf{R}}
\newcommand{\G}{\mathsf{G}}

\newcommand{\ara}{\mathsf{a}}
\newcommand{\arb}{\mathsf{b}}
\newcommand{\arc}{\mathsf{c}}
\newcommand{\ard}{\mathsf{d}}
\newcommand{\CopyB}{\mathsf{copyB}}
\newcommand{\CopyC}{\mathsf{copyC}}

\newcommand{\St}{\mathsf{S}}

\newcommand{\Pt}{\mathsf{P}}

\newcommand{\Array}{\mathsf{array}}

\newcommand{\Merge}{\mathsf{merge}}

\newcommand{\Mersort}{\mathsf{msort}}
\newcommand{\Map}{\mathsf{map}}
\newcommand{\Fib}{\mathsf{f}}

\newcommand{\Fun}{\mathsf{f}}

\begin{document}
\begin{center}
\textbf{\large{Globality and Regions}}\medskip\\
Héctor Gramaglia\footnote{hector.gramaglia@unc.edu.ar}\\
FAMAF, Facultad de Matemática, Astronomía, Física y Computación\\
Universidad Nacional de Córdoba\\
CIEM, Centro de Investigación y Estudios de Matemática
\end{center}

\noindent\textbf{Abstract:}\textit{We obtain a characterization of global variables by unifying abstraction with region abstraction in a region-based language. More precisely,  in a previous work a language called \textsf{global} was presented, whose virtue is to provide a conceptually clear way  of introducing imperative operations in a functional language. Memory safety  is provided by the concept of \textsf{linear protection}, which connects the global system to a linear one. In this paper we show that the concept of global variable provided by the global language arises from the Tofte and Talping's region language through the unification of abstraction and region abstraction.}

\section{Introducción}
\label{introduccion}

Most type-safe languages rely upon a garbage collector to reclaim storage safely.
Region-based memory management presents an alternative to avoiding garbage collection. In this approach, values are allocated on the heap in individually managed regions, and the allocation and deallocation process is performed explicitly.
Region-based memory management employs effect type systems to ensure that unallocated or deallocated regions cannot be accessed at runtime. Tofte and Talping \cite{tofte} present a Region Typed Language (RTL), whose type system guarantees safe memory management (see also \cite{henglein} for another presentation). The judgment $\Gamma\vdash\e:^\varphi\T$  enriches traditional type judgment with a effect $\varphi$, which captures the regions that must be active (i.e. allocated) for the expression to be evaluated without memory problems. The types and effects discipline was developed by Gifford and Lucassen \cite{gifford}, and refined by Jouvelot, Talpin and Toften \cite{jouvelot,jouvelottofte,tofte}.

On the other hand, in \cite{gramagliawlg} we present a language called \textit{global}, which easily introduces imperative components into an applicative program, and which has the virtue of establishing a correspondence between memory deallocation action of update in-place and  weak-linear evaluation: the substructural property granted by the weak-linear type system guarantees the safe introduction of imperative elements.

The objective of this work is to show that, although global evaluation (with linear protection) and region evaluation differ in the mechanism for reclaiming storage, both are closely related: global evaluation takes the concept of global variable from the unification of lambda abstraction and region abstraction. In this way, the in-place update can be modeled through collapsed regions\footnote{When a new value enters the region, the existing one is deleted.} determined by  global variables. To facilitate this task, we will use a language inspired by a fragment of RTL that focuses on the abstraction of regions $\lambda \rho.\mathsf{u}$ as a fundamental device for memory management. Following the line of work in \cite{gramagliawlg} for weak-linear systems, we restrict the form that the expression $\e$ can take in the phrase $\e\ \At\ \sigma$. Much of the power of RTL is resigned, as we are only interested in the annotation of regions, which will allow us to understand the global system in terms of the system of regions.

We attempt to position global language as a bridge between two well-known mechanisms for reasoning about states (linear typing and region-based typing) and imperative language.

\section{A fragment of a RTL}
\label{unfragmentodeRTL}

The language $\lambda[\Sigma^\sigma]$ that we present now is strongly based on a fragment of the language of regions given by Tofte and Talping in \cite{tofte}. Some notation and the type system are taken from \cite{henglein}. This language enriches a functional language with identifiers that denote  lexically-scoped regions, which determine a particular structure for the store, and allow each region to be managed independently

A central aspect of the region-based system is that well-typed programs are memory safe. The phrase $\New\ \rho .\e$ determines the lexical scope for a new region, and its semantics assigns a new region (say $\vr$) to evaluate $\e$. When the evaluation ends, the  region  $\vr$ is deallocated\footnote{Although this aspect has no parallel with memory deallocation in global evaluation (which is linked to linear evaluation), we incorporate this mechanism because it is an essential part of RTL.}, giving rise to a stack-oriented memory management discipline: the most recently allocated region is deallocated first.

The syntax of $\lambda[\Sigma^\sigma]$ is presented in Figure 1. The language is built from a  signature $\Sigma$.  Here $\arc$  (resp. $\op$) represents a 0-ary ($n$-ary) operator on $\Sigma$. The phrases $\x,\vr,\rho$ denote pairwise disjoint infinite families.

The decision to allow an explicit store in a program  $P=(\St,\e)$ is due to the particularities of the global language (section \ref{lenguajeglobal}), which aims to adequately represent global variables. For example, if $\St\ \!\x=\va$, then the program $P=(\St,\x)$ plays the role of closure, in the terms of \cite{tofte} (recursive closure if $\St\ \!\x=\lambda\q.\lambda\p.\e$ and $\x$ is free in $\e$). Evaluating a program of the form $(\St,\e)$ will require the definition of a region-context $\R$ that adequately represents the regions already present in the store $\St$ (see figure 6 of section  \ref{sistemadetiposderegiones}).

The language presents some differences with the RTL presented in \cite{tofte} (beyond the suppression of type polymorphism) that aim to draw a correspondence with the global language $\lambda[\Sigma^g]$ (section \ref{lenguajeglobal}), which in turn is strongly conditioned by its correspondence with the weak-linear language defined in \cite{gramagliawlg}. We adopt the notation  $\New\ \rho.\e$ from  \cite{henglein} as a replacement for the  $letregion$ construct. On the other hand, as we have already noted, we  concentrate the attributes necessary for managing regions in the signature and abstractions, which forces us to suppress  product-type values from this management regime\footnote{The fact that tuples are not storable values forces the use of patterns in lambda abstraction. This aspect, although it introduces some inconvenience in handling the syntax, is fundamental in \cite{gramagliawlg} to reduce the bureaucracy of the syntax in the substructural language.}.

\begin{center}
\begin{tabular}{|c |}
\hline
$
\begin{array}{ll}
\begin{array}{llll}
\vr & \in \ \textsf{Region Name}\medskip\\
\rho & \in \ \textsf{Region Variable}\medskip\\
\q & ::= \  \rho \qquad\qquad  \textsf{Region  pattern}\\
&  \qquad [\rho,...,\rho]
\medskip\\
\sigma & ::= \  \rho\ |\ \vr \qquad\qquad\qquad\ \ \  \textsf{Place}\medskip\\
\varphi & ::= \  \{\sigma,...,\sigma\}\qquad\qquad\  \textsf{Effects} \medskip\\
\p & ::= \  \x \qquad\qquad\qquad\quad\ \textsf{Patterns} \\
&  \qquad \La \p,...,\p \Ra \medskip\\
\V & ::= \ (\B,\sigma)\qquad\qquad \mathsf{Value\  types}\\
  & \qquad     (\Pi\q.\!\T\rightarrow^\varphi \T,\sigma)\medskip\\
  \T& ::=  \ \V \qquad\qquad \textsf{Type expression} \\
  &    \qquad \La \T,...,\T \Ra \smallskip\\

\end{array}
&
\begin{array}{rlllll}
\e& ::= &\x\qquad \qquad\quad \textsf{Expressions}                  \\
  &     & \op( \e,...,\e)\ \At\ \sigma     \\
   &    & \lambda\q.\lambda \p\!:\!\!\T.\ \!\e \ \At\ \sigma\\
  &     & \La \e,...,\e\Ra \\
  &     & \x\ \q\ \e \\
  &     & \If\ \e\ \Then\  \e\ \Else\ \e\\
   &    & \Let\ \p\equiv \e\ \In\ \e\\
  &     & \New \ \sigma.\ \e  \smallskip\\
\St & ::= & \emptyset\qquad\qquad\qquad\quad\ \ \textsf{Store}\\
 & &    \x=\va,\ \St\smallskip\\
P & ::= & (\St,\e)\qquad\qquad\ \  \textsf{Program}\medskip\\
\Gamma    & ::= & []\qquad\qquad\ \mathsf{Type\  context}\\
& & \Gamma,\ \x:\V\\
\end{array}
\end{array}
$\\
\hline
\end{tabular}
\tiny Figure 1: Syntax of $\lambda[\Sigma^\sigma]$
\end{center}

We reiterate that the program $\Pt$ defined by $(\x\!\!=\!\!\lambda\q.\lambda\p.\e,\! \ \e')$ (with $\x\in FV\ \!\e$) fulfills the role of the phrase $\Letrec\ \!\x\equiv \!\lambda\q.\lambda \p\!:\!\!\T.\ \!\e\ \In\ \e'$, thus providing a way to simultaneously incorporate recursion and region polymorphic recursion into the language (see section \ref{sistemadetiposderegiones} and \cite{henglein}).

\subsection{Small-step semantic for $\lambda[\Sigma^\sigma]$}
\label{semanticasmallstepparaL1sigmar}

\textit{Evaluation context} $\e[]$, \textit{Region context} and the \textit{context rule} are defined  in Figure 2. To define small-step semantics we will use context-based semantics, whose distinctive characteristic is the explicit management of the store $\St$, for which we assume that no variables are repeated, and that when extending it, a new variable is used, supplied by $new\ \St$.

\begin{center}
\begin{tabular}{|c |}
\hline
$
\begin{array}{l}
\begin{array}{lll}
(\R_0,\St_0,\e_0) \rightarrow_\beta (\R_1,\St_1, \e_1)\\
\overline{(\R_0,\St_0,\e[\e_0]) \rightarrow (\R_1,\St_1,\e[\e_1])}\bigskip\\
\R  \ ::=\   \emptyset \qquad\qquad\qquad\ \
\textsf{Region}\\
\qquad\quad     \vr\mapsto[\x,...,\x]
, \R \quad  \textsf{Context}\\
\end{array}
\begin{array}{lllllll}
\e[] &::=  & []  \qquad\qquad\qquad \qquad  \textsf{E-Contexts}\\
 &    & \op(\x,...,\x,\e[],...,\e)\ \At\ \sigma\\
&     &  \La \p,...,\p,\e[],..,\e\Ra\\
     &     & \x\ \q\  \e[]\\
     &    &\If\ \e[]\ \Then\ \e\ \Else \ \e\\
     &     &\Let\ \p\equiv\e[]\ \In\ \e\\
     &     &\New\ \sigma.\ \e[]\\
\end{array}\\
\end{array}
$\\
\hline
\end{tabular}
\tiny Figura 2:  Context rule, Region context and Evaluation contexts.
\end{center}

We establish some differences with respect to the semantics defined in \cite{tofte}. On the one hand, we will use small-step semantics where phrases of the form $(\R,\St,\p)$ will be the canonical forms (\textit{values} of \cite{tofte}). In particular, a phrases of the form $(\R,\St,\x)$ where $\St\ \!\x=\lambda\q.\lambda\p.\e$ and $\x$ is free in $\e$, will be the equivalent of a recursive closure. On the other hand we will retain the store (\textit{environment} of \cite{tofte}) in its traditional sense, and we will use a \textit{region context} $\R$ for region management, which has a different meaning from the one given in \cite{tofte}. In this article, $\R$ associates regions with memory locations.

By $\R\sim\n$ we denote the removal of the region $\vr$ from the domain of $\R$ (i.e., $\R|_{dom\R-\{\vr\}}$). On the other hand, if $A$ is a set of variables, then $\St\sim A$ will denote $\St|_{dom\St-A}$. Furthermore, we will use the notation $\R+(\vr,\x)$ as shorthand for $\R+\{\vr\mapsto(\R\ \!\vr\!+\!\!\!+\![\x])\}$, that is, the operation that adds the variable $\x$ to the end of the list $\R\ \!\vr$. Finally, we have the operation $\New\vr\ \R$ that enables a new region, that is, if $\vr=\New\vr\ \R$ then $\vr\notin dom\ \R$.

By  $[\p\mapsto\p']$ we extend the substitution\footnote{$\x_1\mapsto\y_1$ denotes the identity map modified in the variable $ \x_1$, where it takes the value $\y_1$. } $\x\mapsto \y$ to patterns. Such extension is given by the conditions: $[\x\mapsto\y]= \x\mapsto \y$, $[\ep\mapsto \p]= []$ and

\medskip
$
\begin{array}{rlllll}
\left[\La \p_1,...,.\p_n\Ra\mapsto \La \p'_1,...,.\p'_n\Ra\right]  & = &
\left[\p_1\mapsto \p'_1\right],..., \left[\p_n\mapsto \p'_n\right]
\end{array}
$

\medskip

Terminal configurations will be triples of the form $(\R,\St,\p)$. In the initial configuration $(\R,\St,\e)$ the free variables of $\e$ will be assigned a region through the context $\R$ (see rule \textsf{sba} in Figure 6). In Figure 3 the rules of the small-step semantics are given.

\begin{center}
\begin{tabular}{|c |}
\hline
$
\begin{array}{llll}
(\mathsf{ela}) & (\R,\St,\lambda\q.\lambda\p\!:\!\!\T.\ \!\e\ \At\ \vr) \rightarrow_\beta\\
&\qquad \qquad \qquad (\R\!+\!(\vr,\x),\St,\x=\lambda\q.\lambda\p.\e,\x) \smallskip\\
\mathsf{(eop)} & (\R,\St,\op(\x_1,...,\x_k)\ \At\ \vr) \rightarrow_\beta
&
(\w=\op(\w_1,...,\w_n)\\
&\qquad\qquad \qquad \qquad(\R\!+\!(\vr,\x),\St,\x=\w,\x)& \ \St\ \x_i= \w_i,\ \x=\New\ \ \St)\smallskip\\
(\mathsf{eif}) & (\R,\St, \If\ \x\ \Then\ \e_0\ \Else\ \e_1)\rightarrow_\beta(\R,\St,\e_0) & (\St\x= \True)\smallskip\\
 & (\R,\St, \If\ \x\ \Then\ \e_0\ \Else\ \e_1)\rightarrow_\beta(\R,\St,\e_1) & (\St\x= \False)\smallskip\\
(\mathsf{ele}) & (\R,\St, \Let\ \p\equiv\ \p'\ \In\  \e) \rightarrow_\beta (\R,\St, [\p\mapsto\p']\e) &\smallskip\\
(\mathsf{eap}) & (\R,\St, \f\ \q'\ \p') \rightarrow_\beta (\R,\St, [\q\mapsto\q'][\p\mapsto\p']\e) & (\St \f=\ \lambda\q.\lambda \p:\T.\e)\smallskip\\
\mathsf{(ene)} & (\R,\St,\New\ \rho.\ \e) \rightarrow_\beta(\R,\vr\mapsto[],\\
& \qquad\qquad\qquad\qquad\quad \St,\New\ \vr.\ [\rho\mapsto\vr]\e)&(\vr=\New\vr\ \R)\smallskip\\
\mathsf{(ede)} & (\R,\St,\New\ \vr.\p) \rightarrow_\beta(\R\sim\vr,\St\sim(\R\ \!\vr),\p)&\smallskip\\
\end{array}
$\\
\hline
\end{tabular}
\tiny Figura 3: Small-step semantic of $\lambda[\Sigma^\sigma]$
\end{center}

\subsection{Region Type System}
\label{sistemadetiposderegiones}

An essential feature of Tofte
and Talpin's region language
compared to other region language (such as \cite{hanson}, \cite{ross}, \cite{schwartz}) is the presence of a type system that guarantees that programs are memory-safe. In the $\Gamma\vdash\e:^\varphi \T$ judgment, the $\varphi$ effect captures the regions that must be active (i.e., allocated) for the term to be evaluated without memory problems.

The type system presented here is based on the system presented by Henglein, Makholm and Niss in \cite{henglein}, which in comparison to Tofte and Talpin's system \cite{tofte},
move the effect enlargement upwards in the derivation tree, so that the axioms are responsible for introducing proper effects.

For handling patterns we will need the following notation. We use $[\p:\T]$ to denote the phrase of type $\Gamma$ which consists of flattening the pattern $\p$ and the type $\T$. That is, we define $\left[\x:\V\right] =  \x:\V$,   and

\medskip
$
\begin{array}{lll}
\left[\La\p_i,...,\p_n\Ra : \La\T_1,...,\T_n\Ra\right] & = & \left[\p_1 : \T_1\right],...,\left[\p_n : \T_n\right]\\
\end{array}
$

\medskip

The rules of the type system are given in Figures 5 and 6. In rule \textsf{bop} we assume that $\op$ is a basic operator symbol of $\Sigma$. In rule \textsf{new}, function $\mathsf{Eff}(\Gamma\!,\!\T)$ denotes the set of places that occur freely in $\Gamma$ and $\T$.

\begin{center}
\begin{tabular}{|c |}
\hline
$
\begin{array}{ll}
(\mathsf{var})
\begin{array}{l}
\quad \\
\quad \\ 
\overline{\Gamma_1, \x :\V,\Gamma_2\vdash \x :^{\varphi}\V}\\
\end{array}
&
(\mathsf{tup})
\begin{array}{l}
\quad\Gamma\vdash \e_i:^\varphi\T_i\\
\overline{\Gamma\vdash\La\e_1,...,\e_n\Ra:^\varphi\La\T_1,...,\T_n\Ra}\\
\end{array}
\bigskip\\
(\mathsf{new})
\begin{array}{l}
\quad\sigma\notin \mathsf{Eff} (\Gamma,\T)\\
\quad\Gamma\vdash \e:^{\varphi\cup\{\!\sigma\!\}}\T\\
\overline{\Gamma\vdash \New\ \sigma.  \ \e:^{\varphi}\T}\\
\end{array}
&
(\mathsf{con})
\begin{array}{l}
\quad\Gamma\vdash \e:^\varphi(\Bool,\sigma)\\
\quad\Gamma\vdash \e_i:^\varphi\T\\
 \overline{\Gamma\vdash\If\ \e\ \Then\ \e_1\ \Else\ \e_2:^\varphi\T}\\
\end{array}
\bigskip\\
(\mathsf{let})
\begin{array}{l}
\quad\Gamma\vdash \e:^\varphi\T\\
\quad\left[\p\!:\!\!\T\right] \vdash \p:^{\varphi}\T\\
\quad\Gamma,\left[ \p\!:\!\!\T\right]\vdash \e':^\varphi\T'\\
\overline{ \Gamma\vdash \Let\  \p\equiv\e\ \In\ \e':^\varphi\T'}\\
\end{array}
&
(\mathsf{app})
\begin{array}{l}
\quad [\q\mapsto\f]\varphi'\subseteq\varphi\\
\quad\Gamma\ \h=\Pi\q\!:\!\D\rightarrow^{\varphi'}\T\\
\underline{\quad\Gamma\vdash \e:^\varphi[\q\mapsto\f]\D\qquad}\\
\quad\Gamma\vdash\ \h\ \f\ \e:^\varphi[\q\mapsto\f]\T\\
\end{array}
\bigskip\\
(\mathsf{lam})
\begin{array}{l}
\quad\sigma\in\varphi,\ \
\q\cap \mathsf{Eff}\ \!\Gamma=\emptyset\\
\quad\left[ \p\!:\!\D\right] \vdash \p:^{\varphi}\D\\
\quad\Gamma,[\p\!:\!\D]\vdash \e:^{\varphi'}\T\\
\overline{\Gamma\vdash \lambda\q.\lambda\p\!:\!\D.\ \!\e\ \At\ \sigma:^\varphi}\\
\qquad \qquad\quad(\Pi\q\!:\!\D\rightarrow^{\varphi'}\!\T,\sigma
)\smallskip\\
\end{array}
&
(\mathsf{bop})
\begin{array}{l}
\quad\op:(\B_1,...,\B_n)\!\rightarrow\!\B\\
\quad\sigma\in\varphi\\
\quad\Gamma_i\vdash \e_k:^{\varphi}(\B_k,\sigma_k)\\
\overline{\Gamma\vdash \op(\e_1,...\e_n )\ \At\ \sigma:^{\varphi}(\B,\sigma)\smallskip}\\
\end{array}
\\
\end{array}
$\\
\hline
\end{tabular}\\
\tiny Figura 5: $\Gamma\vdash\e:\T$
\end{center}

\begin{center}
\begin{tabular}{|c |}
\hline
$
\begin{array}{l}
(\mathsf{sem})
\begin{array}{c}
\\
\underline{\ \varphi\subseteq dom\ \!\R\ }\\
\R \vdash []:^\varphi[]\\
\end{array}
\quad\qquad\qquad\
(\mathsf{sba})
\begin{array}{c}
\\
\R\vdash\St:^\varphi \Gamma\quad\n\in\varphi\quad(\op:\B)\in\Sigma\\
\overline{\R+(\n,\x)\vdash \St,\x=\op:^{\varphi}(\Gamma,\x\!:\!(\B,\n))}\\
\end{array}
\medskip\\
(\mathsf{sfu})
\begin{array}{c}
\underline{\R\vdash\St\!:^\varphi\! \Gamma \quad  \n\!\in\!\varphi\quad\left[\p\!:\!\D\right]\vdash \p\!:^\varphi\!\D \quad \Gamma,\f\!:(\!\Pi\q\!:\!\D\!\rightarrow^{\varphi'}\!\!\T,\n), \!\left[\p\!:\!\D\right]\vdash^{\varphi'}\! \e\!:\!\!\T}\\
\R+(\n,\f)\vdash \St,\f=\lambda \q.\lambda \p.\e :^\varphi(\Gamma,\f:(\Pi\q:\!\D\rightarrow^{\varphi'}\!\T,\n))\\
\\
\end{array}
\end{array}
$\\
\hline
\end{tabular}\\
\tiny Figura 6: $\vdash\St:\Gamma$
\end{center}

Finally, the relation $\R\vdash^\varphi(\St,\e) $ is defined by the rule:

\medskip
\qquad\qquad\qquad\qquad$
\begin{array}{c}
\underline{\R\vdash\St:^\varphi \Gamma\qquad \Gamma\vdash \e:^\varphi\T}\\
\R\vdash^\varphi (\St,\e)\\
\end{array}
$
\medskip

For the examples we will use the notation $\op^\sigma(\e_1,...,\e_n)$ as an abbreviation for $\op(\e_1,...,\e_n)\ \At\ \sigma$. We also write $\e_1\ \op^\sigma\ \e_2$ if $\op$ is a binary operation.

In \cite{tofte} the authors experiment with the following version of fibonacci (see also \cite{henglein}), which we will call $fib_1$, observing that it requires few memory resources, although it performs a large number of region allocations and deallocations\footnote{The examples were developed and tested in the Haskell prototype which can be found at https://github.com/hgramaglia/GloReg.}. In the context $\f:(\Pi[\rho,\kappa].(\Int,\kappa)\rightarrow(\Int,\rho),3)$, the program has type $(\Int,1)$.

\medskip
\noindent$fib_1$

\noindent\begin{tabular}{|c |c |}
\hline
$
\begin{array}{l}
\Fib\ = \ \lambda [\rho,\kappa].\lambda  \x .\ \New\ \beta.\\
\qquad\qquad\If\ \x <^\beta\ 2^\beta \ \Then\ 1^{\rho}\\
\qquad \qquad\Else\
\New\ \rho_1.\\
\qquad\qquad\quad\ (\New\ \kappa_1.\ \Fib \Lc\rho_1,\kappa_1\Rc\ ((-1)^{\kappa_1}\ \x)) \ +^\rho\ (\New\ \kappa_2.\ \Fib \Lc\rho_1,\kappa_2\Rc\ ((-2)^{\kappa_2}\ \x))\\
\Fib \Lc 1,2\Rc\ n^{2}\\
\end{array}
$\\
\hline
\end{tabular}\\

\medskip

We note the importance in the example shown of the fact that recursive calls can use actual parameters other than the formal parameters, which prevents the generation of long-lived regions. This is known in the literature as \textit{region polymorphic recursion}. Although this is an important phenomenon in relation to RTL, we must clarify that it is contrary to the region management that arises from globalization, which promotes long-lived regions.

\section{Global language}
\label{lenguajeglobal}

The language defined in \cite{gramagliawlg} presents some differences\footnote{We make explicit as an effect the set of global variables in the function types and in the relation $\vdash$.} with the one we are now introducing, which aims to highlight its connection with the language of regions. The language defined in \cite{gramagliawlg} is built from a qualified signature $\Sigma^\g$. The main function of this signature is to define the concept of \textit{substructural protection}, which guarantees the correctness of the global (imperative) version. In this work it will not be necessary to define the concept of qualified signature, since in order to relate the global system with the system of regions it will be convenient to make the global qualifier explicit in each occurrence of a signature operation (phrase $\op(...)\ \At\ \g$). We keep the superscript $\g$ in the name $\lambda[\Sigma^\g]$\footnote{In \cite{gramagliawlg} the name $L^1[\Sigma^\g]$ was used for a more reduced language. The languajes $L^1[\Sigma^\q],L^1[\Sigma^\g]$ defined in this work do not incorporate functions or tuples in the special memory management regimes (linearity and globality). Tuples are not storable values, which forces the use of patterns in lambda abstraction. This feature is maintained in $\lambda[\Sigma^\sigma]$ and $\lambda[\Sigma^\g]$.} that we will use for the global language, and in this article we will put it in relation to the region language $\lambda[\Sigma^\sigma]$ (section \ref{unfragmentodeRTL}). The names are due to the fact that, following the line of \cite{gramagliawlg}, we concentrate the studied attributes (substructurality, in-place update or regions) in the signature operators and abstraction. The superscripts $\g$ (with $\g=\lo,\x$) and $\sigma$ (with $\sigma=\n,\rho$) indicate the qualifiers with which the language is enriched using  $\At$ annotation. In this line, the notation $(\B,\g)$ of the global language corresponds to the notation $\g\ \B$ of \cite{gramagliawlg} (which comes from the substructural systems \cite{walker}), while the notation $\op(...)\ \At\ \g$ corresponds to $\op^\theta(...)$ of the cited article, where $\theta$ is of the form $(..)\rightarrow\ \g\ \B$.

Regardless of the language ($\lambda[\Sigma^\g]$ or $\lambda[\Sigma^\sigma]$), and consequently, regardless of the nature of the phrase $q$ in the phrases $(\B,q)$ and $\op(...)\ \At\ q$, we will refer to it as a qualifier, inheriting the notation specific to the substructural systems.

In the language presented in \cite{gramagliawlg}, global qualification of operators establishes the difference between a functional program (trivial qualification: all $\lo$) and a global program, which incorporates imperative elements through non-trivial qualification. Qualifiers are used to enrich the types of the $\Sigma$ operators, because the correctness of the introduction of imperative elements is obtained by relating $\Sigma^\g$ with the signature that grants substructurality\footnote{The substructural language uses  qualifiers $\qu,\ql$ to delimit the life cycle of a value through its type.} to the program.

The abstract syntax of the language $\lambda[\Sigma^\g]$ is shown in Figures 7 and 8. The abstract phrase $\x$ represents an infinite set of variables. A program will be a pair $P=(\St,\e)$. As will be seen in section  \ref{semanticassmallstep}, the fixed-point operator is implicit: function definitions in $\St$ will be considered recursive\footnote{That is, if $\St\ \!\x=\lambda\p.\e$ and $\x$ is free in $\lambda\p.\e$. In the terminology of \cite{tofte}, $\x=\lambda\p.\e$ is a recursive closure. } as long as the variable being defined is free in the body of the definition.

\begin{center}
\!\!\begin{tabular}{|c |}
\hline
$

\begin{array}{ll}
\begin{array}{llll}
\g ::= & \x &\!\!\!\!\mathsf{global\  qualifiers}\\
  & \lo\smallskip\\
\p  ::= & \x &\!\!\!\!\mathsf{patterns}\\
  & \La \p,...,\p \Ra\smallskip\\
\varphi  ::= & \{\x,...,\x\} &\!\!\!\!\mathsf{global\  effects}\smallskip\\
\B  ::= & \Int &\!\!\!\!\mathsf{basic\ pretype}\\
        & \Bool\\
        & \Array\ ...&\smallskip\\
\end{array}
&
\begin{array}{llll}
 \V ::= (\B,\g) &\mathsf{value\  types}\smallskip\\
\qquad\ \ \!(\Pi\p.\!\T\!\rightarrow^\varphi\!\T\!,\g)\smallskip\\
\T  ::= \V & \textsf{type expression}\smallskip\\
        \qquad\ \ \La \T,...,\T \Ra\smallskip\\
\Gamma    ::=  [] & \!\!\!\!\mathsf{type\  context}\smallskip\\
\qquad\ \  \Gamma,\ \x:\V\\
          \\
\\
\end{array}
\end{array}
$\\
\hline
\end{tabular}
\tiny Figura 7: Global Qualifiers and globally qualified signature\end{center}

\begin{center}
\begin{tabular}{|c |}
\hline
$
\begin{array}{ll}
\begin{array}{lll}
\\
\va & ::= & \op\qquad\qquad\quad(\op\in\Sigma\textsf{ of arity } 0)\\
    &     & \lambda\p.\e\medskip\\
\St & ::= & \emptyset\\
    &     & \x=\va,\ \St\\
    \\
    \\
\end{array}
&
\begin{array}{rlllll}
\e& ::= & \x                    \\
  &     & \op( \e,...,\e)\ \At\ \g     \\
  &     & \lambda\p. \e \ \At\ \g\\
  &     & \La \e,...,\e\Ra \\
  &     & \x\ \e \\
  &     & \If\ \e\ \Then\  \e\ \Else\ \e \qquad\\
   &    & \Let\ \p\equiv \e\ \In\ \e \\
\end{array}
\end{array}
$\\
\hline
\end{tabular}
\tiny Figure 8: Syntax of $\lambda[\Sigma^\g]$
\end{center}

Note that we also use $\Gamma,\T$ to denote global contexts and global types. Depending on the signature involved ($\Sigma^\sigma$ or $\Sigma^\g$) we can distinguish whether we are talking about region or global typing.

The \textit{global} modality, given by the qualified type $(\B,\x)$, will allow us to both add the in-place update and systematize the concept of a global variable. For the \textit{local} modality (pure applicative fragment), we will use the notation $\B$ and $\op(\e_1,...,\e_n)$ as abbreviations for $(\B,\lo)$ and $\op(\e_1,...,\e_n)\ \At\ \lo$, respectively.

We already mentioned that the $\lo$ qualifier determines the purely functional part of the language, in contrast to the variable qualifier, which introduces imperative elements. In relation to the region language, the $\lo$ qualifier refers to an indeterminate region (we will use region 0 for this part of the heap), while the variable qualifier identifies a precise region. We discuss this relationship in the \ref{globalityandregions} section.

The syntax of $\lambda[\Sigma^\g]$ shows a suggestive similarity to the syntax of $\lambda[\Sigma^\sigma]$, except for the phrase $\New\ \sigma.\e$. In the global memory reclamation regime (which models in-place update) there is nothing similar to the memory block management proposed by RTL, while that regime is connected to the memory reclamation mechanism of linear languages. As we have already stated, the connection between $\lambda[\Sigma^\g]$ and $\lambda[\Sigma^\sigma]$ arises from the unification in the RTL of the two abstraction concepts present in the language. It is surprising (we could say uncomfortably surprising) that we are showing a relationship between languages that model imperative operations, but their connection is not given in these attributes. The relationship between these languages will take a strange form: global variables will be identified with ``collapsed'' regions.

\subsection{Small-step semantic}
\label{semanticassmallstep}

\textit{Evaluation context} $\e[]$ and the \textit{context rule} are defined in Figure 9.

Terminal configurations will be pairs of the form $(\St,\p)$. We take the same program $(\St,\e)$ as the initial configuration. In Figure 10 the rules of the small-step semantics are given.

\begin{center}
\begin{tabular}{|c |}
\hline
$
\begin{array}{l}
\begin{array}{lllllll}
(\St_0,\e_0) \rightarrow_\beta (\St_1, \e_1)\\
\overline{(\St_0,\e[\e_0]) \rightarrow (\St_1,\e[\e_1])}\\
\end{array}
\begin{array}{lllllll}
\e[] &::=  & []  \qquad\qquad\qquad\qquad\qquad\ \  \textsf{E-Contexts}\\
 &    & \op(\x,...,\x,\e[],...,\e)\ \At\ \g\\
&     &  \La \p,...,\p,\e[],..,\e\Ra\\
     &     & \x\  \e[]\\
     &    &\If\ \e[]\ \Then\ \e\ \Else \ \e\\
     &     &\Let\ \p\equiv\e[]\ \In\ \e\\
\end{array}\\
\end{array}
$\\
\hline
\end{tabular}
\tiny Figura 9: Evaluation contexts and contexts rule.
\end{center}

\begin{center}
\begin{tabular}{|c |}
\hline
$
\begin{array}{lll}
\mathsf{(ela)} & (\St,\lambda\p.\e\ \At\ \x) \rightarrow_\beta(\St\sim\x,\x=\lambda\p.\e,\x)&\smallskip\\
& (\St,\lambda\p.\e) \rightarrow_\beta(\St,\x=\lambda\p.\e,\x)&(\x=\New\ \St)\smallskip\\
\mathsf{(eop)} & (\St,\op(\x_1,...,\x_n)\ \At\ \x) \rightarrow_\beta(\St\sim\x,\x=\w,\ \x) & (\St\x_i= \w_i,\\
&  & \ \w=\op(\w_1,...,\w_n))\smallskip\\
 & (\St,\op(\x_1,...,\x_n)) \rightarrow_\beta(\St,\x=\w,\x) & (\x=\New\ \St,\ \St\x_i= \w_i,\\ & & \w=\op(\w_1,...,\w_n))\smallskip\\
(\mathsf{eif}) & (\St, \If\ \x\ \Then\ \e_0\ \Else\ \e_1)\rightarrow_\beta(\St,\e_0) & (\St\x= \True)\smallskip\\
 & (\St, \If\ \x\ \Then\ \e_0\ \Else\ \e_1)\rightarrow_\beta(\St,\e_1) & (\St\x= \False)\smallskip\\
(\mathsf{ele}) & (\St, \Let\ \p\equiv\ \p'\ \In\  \e) \rightarrow_\beta (\St, [\p\mapsto\p']\e) &\smallskip\\
(\mathsf{eap}) & (\St, \h\ \p') \rightarrow_\beta (\St, [\p\mapsto\p']\e) & (\St\ \! \h=\ \lambda \p.\e)\smallskip\\
\end{array}
$\\
\hline
\end{tabular}
\tiny Figura 10: Small-step semantic
\end{center}

\subsection{Global type system}
\label{sistemadetipoglobal}
The type system of $\lambda[\Sigma^\g]$ is intended to capture the imperative nature of the program. This will be reflected in the map  $\mathcal{I}$ that we will define in section \ref{formaimperativa}.
It contains elements of a dependent type system, but can be fully understood from the system defined in section  \ref{sistemadetiposderegiones}. The type system in no way guarantees safe memory management. In \cite{gramagliawlg} this property is granted (for a fragment of $\lambda[\Sigma^\g]$) by substructural  protection.

Before giving the rules of the global type system, let's give some technical definitions.
The $\lambda[\Sigma^\g]$ types carry information about the store. The amount of information a type has is compared using the relation $\T\leq\T'$.
To define this relation, we define the poset $X$ as the lifting of the set of variables (with the flat order) with the smallest element $\lo$. The set of patterns $\p$ is embedded in the partially ordered set $Pat$ formed by the direct sum of all possible combinations of direct products of $X$\footnote{For example $\La \x,\La\z,\lo\Ra\Ra$ is an element of $\La X,\La X, X\Ra\Ra$, and $\La \x,\La\z,\lo\Ra\Ra\leq\La \x,\La\z,\w\Ra\Ra$ is verified (this last element is a supremum).}.

The relation $\T\leq\T'$ is determined by the condition $\T\leq\T' \Leftrightarrow\p_\T\leq\p_{\T'}$, where the map $\T\rightarrow\p_\T$ is defined by the following conditions:

\medskip

$
\begin{array}{rlll}
\p_{(\B,\g)}& = &\g& \\
\p_{(\Pi\p:\T_1.\!\T_2,\g)}& = &\g\\
\p_{\La\T_1,...,\T_n\Ra}& = & \La\p_{\T_1},...,\p_{\T_n}\Ra\\
\end{array}
$

\medskip

In our type system, the function type $\T\rightarrow \T'$ is replaced by the type $(\Pi\p.\!\T\rightarrow^\varphi\!\!\T',\g)$, where the pattern $\p$ is a transmitter of information about the store that carries the argument of type $\T$. Then the type $(\Pi\p.\!\T\rightarrow^\varphi\!\!\T',\g)$ will only make sense when $\p$ can faithfully carry the information about the store that has the type $\T$. This condition is formalized by the property $\p_\T\leq\p$. This means that, for example, the type $(\Pi \x.(\z,\Int)\rightarrow^\varphi\!\T)$ is meaningless to us.

In the application $\f\ \e$, with $\St\f=\lambda\p.\e_0$ and $\Gamma\f=(\Pi\p.\!\T\rightarrow^\varphi\!\!\T',\g)$, the pattern $\p$ (by a substitution\footnote{Trivially defined.}) will transmit to $\T'$ the information about the store carried by $\e$. This information to be transmitted will take the form of an object of $Pat$. We define $p^\Gamma\e\in Pat$ by the following conditions.
\medskip

\noindent$
\begin{array}{lcll}
p^\Gamma\ \x & = &  \x\\
p^\Gamma\ (\op(\e_1,...,\e_n)\ \At\ \g)  & = & \g \\
p^\Gamma\ (\lambda\p.\e\ \At\ \g) & = &  \g\\
p^\Gamma\ (\If\ \e_1\ \Then\ \e_2\ \Else\ \e_3) & = &p^\Gamma\  \e_2\\
p^\Gamma\ \La\e_1,...,\e_n\Ra & = & \La p^\Gamma\ \e_1,...,p^\Gamma\ \e_n\Ra &\\
p^\Gamma\ (\Let\ \p\equiv\e_1\ \In\ \e_2) & = &[\p\mapsto p^\Gamma\e](p^\Gamma\  \e_2)\\
p^\Gamma\ (\f\ \e) & = &  \p_{[\p\mapsto p^\Gamma\e]\T'}\qquad\qquad\quad (\Gamma\f=( \Pi\p.\!\T\rightarrow^\varphi\T',\g))\\
\end{array}
$

\medskip

We now give the typing rules. The \textsf{loc} rule shown below presents the distinctive character of the type system, and expresses that the type takes from the variable (as a memory location) information about the store. The \textsf{var} rule is a typical typing rule for variables of basic types and functions.

\medskip

$\mathsf{(loc)}
\begin{array}{c}
\Gamma\vdash^\varphi \x :(\Pt,\lo) \\
\overline{\ \Gamma\vdash^\varphi \x :(\Pt,\x)\ }\\
\end{array}
\ \ (\Pt=\B,\ \Pi\p.\!\T\rightarrow^\varphi\!\!\T')
\qquad\mathsf{(var)}
\begin{array}{c}
\\
\overline{\Gamma_{1},\x :\V,\Gamma_{2}\vdash^\varphi \x :\V}\\
\end{array}
$

\medskip

\noindent
Note that \textsf{loc} allows an expression to be typed with several different types, all of them differing in the level of information about global variables that the type carries.
This feature is central to the overall system: a function, operation or constructor can receive data of the type $\T$ that corresponds to its specification, or data of type $\T'$ that carries more information than expected ($\T\leq\T'$). This is made possible by the rule:

\medskip

$\mathsf{(\leq)}
\begin{array}{c}
\underline{\Gamma\vdash^\varphi \e :\T'\quad \T\leq\T'}\\
\Gamma\vdash^\varphi \e :\T\\
\end{array}
$

\medskip

The rest of the rules are given in Figure 11 and 12. Below we define the context operator $\Gamma_1;\Gamma_2$, used in Figure 9. The explicit presence of the global effect in the type of a function is only intended to facilitate the transformation of a global program into a region one.

\begin{center}
\begin{tabular}{|c |}
\hline
$
\begin{array}{ll}
 \\
(\mathsf{bop})
\begin{array}{l}
\quad(\op\!:\!(\B_1,...,\B_n)\rightarrow\B)\in\Sigma\\
\quad\g\in\varphi\\
\quad\Gamma\vdash^\varphi \e_i:(\B_i,\g_i)\\
\overline{\Gamma\vdash^\varphi \op(\e_1,...,\e_n )\ \At\ \g:(\B,\g)}\\
\end{array}
&
(\mathsf{app})
\begin{array}{l}
\quad[\p\mapsto p^\Gamma\e]\varphi'\subseteq\varphi\\
\quad\Gamma\ \f=\Pi\p.\!\T\rightarrow^{\varphi'}\!\T'\\
\quad\Gamma\vdash^\varphi \e: [\p\mapsto p^\Gamma\e]\T \\
\overline{\Gamma\vdash^\varphi \f\ \e:[\p\mapsto p^\Gamma\e]\T'}\\
\end{array}
\medskip\\
(\mathsf{let})
\begin{array}{l}
\quad\Gamma\vdash^\varphi \e:\T\\
\quad[\p\!:\!\!\T]\vdash^\varphi\p: \T\\
\quad\Gamma; [\p\!:\!\! \T]\vdash^\varphi \e':\T'\\
\overline{\Gamma\vdash^\varphi \Let\  \p\!\equiv\!\e\ \In\ \e'\!:\!\T'}\\
\end{array}
&
(\mathsf{lam})
\begin{array}{l}
\quad\g\in\varphi\\
\quad[\p\!:\!\!\T]\vdash^\varphi\p: \T\\
\quad\Gamma; [\p\!:\!\! \T]\vdash^{\varphi'} \e:\T'\quad \\
\overline{\Gamma\vdash \!\lambda\p.\e\ \!\At\ \!\g :\!(\Pi\p.
\!\T\!\!\rightarrow^{\varphi'}\!\T'\!\!,\g)}\\
\end{array}
\medskip\\
(\mathsf{tup})
\begin{array}{l}
\quad\Gamma\vdash^\varphi \e_i:\T_i\\
\overline{\Gamma\vdash^\varphi\La\e_1,...,\e_n\Ra:\La\T_1,...,\T_n\Ra}
\end{array}
&
(\mathsf{con})
\begin{array}{l}
\quad\Gamma\vdash^\varphi \e:\g\ \Bool\\
\quad\Gamma\vdash^\varphi \e_i:\T\\
 \overline{\Gamma\vdash^\varphi \If\ \e\ \Then\ \e_1\ \Else\ \e_2:\T}\\
\end{array}
\medskip
\\
\end{array}
$\\
\hline
\end{tabular}\\
\tiny Figura 11: $\Gamma\vdash\e:\T$
\end{center}

\begin{center}
\begin{tabular}{|c |}
\hline
$
\begin{array}{l}
(\mathsf{sem})
\begin{array}{c}
\\
\underline{\quad\quad\quad\quad\quad}\\
\ \vdash^\varphi [ ]:[]\\
\end{array}
\qquad\
(\mathsf{sba})
\begin{array}{c}
\\
\underline{\ \vdash^\varphi\St : \Gamma\quad\g\in\varphi\quad  (\op:\B)\in\Sigma\ }\\
\vdash^\varphi \St,\x=\op:\Gamma,\x:(\B,\g)\\
\end{array}
\ \ (\g=\lo,\x)
\medskip\\
(\mathsf{sfu})
\begin{array}{c}
\underline{\vdash^\varphi\St: \Gamma \quad \g\in\varphi\quad \left[\p\!:\!\!\T\right]\vdash^\varphi\! \p\!:\!\!\T \quad \Gamma,\f:\!(\Pi\p.\!\T\rightarrow^{\varphi'}\!\T',\g), \left[\p\!:\!\!\T\right]\vdash^\varphi \e\!:\!\T'}\\
\vdash^\varphi \St,\f=\lambda \p.\e: \Gamma,\f:(\Pi\p.\!\T\rightarrow^{\varphi'}\T',\g)\\
\\
\end{array}
\end{array}
$\\
\hline
\end{tabular}\\
\tiny Figura 12: $\vdash\St:\Gamma$
\end{center}

Well-typed  $\lambda[\Sigma^\g]$-programs will be called \textit{global}.

The existence of global variables in the context $\Gamma$ forces a restricted handling of bound variables in $\Let$ statement. Classical type systems use the environment operation $\Gamma_1,\Gamma_2$, where a variable that occurs in both contexts is redefined by $\Gamma_2$ overriding its original definition set in $\Gamma_1$.
In the global system this operation must be restricted, becoming a partially defined operation: a global variable (i.e. of type $(\B,\x)$) that occurs in $\Gamma_1$ cannot be redefined.
In the following definition we use the predicates $\lo(\V)$, which is defined by the following conditions: $\lo(\Pt,\g)=(\g\!\!==\!\!\lo)$ (with $\Pt=\B,\Pi\p.\!\T\!\rightarrow^{\varphi}\!\!\T'$)  and $\gl(\V)=\neg\lo(\V)$.

\medskip

$
\begin{array}{rlll}
\Gamma_1;[] & = & \Gamma_1\\
(\Gamma^1_1,\x:\V_1,\Gamma_1^2);(\x:\V_2,\Gamma_2) & = & (\Gamma^1_1,\x:\V_1,\Gamma_1^2);\Gamma_2 & \textsf{if }\gl(\V_1)\wedge\gl(\V_2)\\
(\Gamma^1_1,\x:\V_1,\Gamma_1^2);(\x:\V_2,\Gamma_2) & = & (\Gamma^1_1,\Gamma_1^2,\x:\V_2);\Gamma_2 & \textsf{if }\lo(\V_1)\\
(\Gamma^1_1,\x:\V_1,\Gamma_1^2);(\x:\V_2,\Gamma_2) & = & (\Gamma^1_1,\Gamma_1^2);\Gamma_2 & \textsf{if }\gl(\V_1)\wedge\lo(\V_2)\\
\end{array}
$

\medskip

The rule for  $\Let$-construction reports the existence of a global variable. Indeed, if $\Gamma \x=(\B,\x)$ and $\x$ occurs in $\p$, then the restriction imposed by the operator $(;\!)$ forces the type $(\B,\x)$ for the data corresponding to $\x$.

We end the section by pointing out that typing is strongly nondeterministic, due to the rule $(\leq)$. One possible implementation is to restrict its application to arguments of function calls (rule \textsf{app})  and basic operators (\textsf{bop}). These modifications are shown below.

\medskip

$\begin{array}{l}
\quad\Gamma\ \f=\Pi\p.\!\T_0\!\rightarrow^\varphi\!\T'\\
\quad\Gamma\vdash \e: \T \\
\quad[\p\mapsto p^\Gamma\e]\T_0\leq\T\\
\overline{\Gamma\vdash \f\ \e:[\p\mapsto p^\Gamma\e]\T'}\\
\end{array}$
\qquad
$\begin{array}{l}
\quad \g_i=p^\Gamma \e_i\\
\quad(\op:(\B_1,..., \B_n)\!\rightarrow\!\B)\in\Sigma\\
\quad\Gamma\vdash \e_i:(\B_i,\g_i)\\
\overline{\Gamma\vdash \op(\e_1,...,\e_n )\ \At\ \g:(\B,\g)\quad}\\
\end{array}
$
\medskip

\noindent In the last rule we force the most informative typing option: if it occurs $\Gamma\vdash\e_i:(\B_i,\g^0_i)$, then $\g_i^0\leq\g_i$ (in $Pat$).

The global system generates globality through a key difference observed in the \textsf{(lam)} rule in the two presented systems: in the global system, the restriction on the effects $\q \cap \mathsf{Eff}\ \!\Gamma=\emptyset$ is eliminated. This decision must be complemented by the operation $(;\!)$ on type contexts, which guarantees consistency in the occurrences of a global variable. This modification to the global system eventually nullifies the effect of the lambda abstraction, which forces us to ask why we should include a global variable in the abstraction pattern. The answer lies in the link between the global system and the linear one: an unrestricted lambda abstraction cannot contain linear objects (see \cite{walker}, rule \textsf{T-ABS} on p. \!10). Therefore, if the update of a global variable is protected by the linearity of the object it denotes, that variable must be part of the abstracted pattern. For this reason, some of the examples shown in section  \ref{globalityandregions} should extend their abstraction patterns to obtain linear protection. We will not undertake this task in order to prioritize the clarity of the link between the global and region systems.

\subsection{Imperative form for global programs}
\label{formaimperativa}

In this section we will show that a well-typed $\lambda[\Sigma^\g]$ program can take an imperative form, which consists of making assignment and global variables explicit.

We give below the definition of $\mathcal{I}\ \!G\ \!(\St,\e)$, where $G$ will represent a set of variables (variables to be globalized). If there exist $\Gamma,\T,\varphi$ such that $\vdash\St:^\varphi\Gamma$ and $\Gamma\vdash\e:^\varphi\T$, then the imperative form of the program $(\St,\e)$ is obtained by $\mathcal{I}\ (Gl\ \Gamma)\ (\St,\e)$, where $Gl\ \Gamma$ is the set of variables that occur in $\bigcup_{\x\in dom\Gamma}FV\ \!(\Gamma\x)$.

In the following definitions, $\pp$ denotes the pattern $\p$ in which the variables of $G$ are replaced by $\ep$, leaving the rest unchanged.

\medskip

\noindent$
\begin{array}{rclll}
\mathcal{I}^\e\ G\ \x & = & \x \\
\mathcal{I}^\e\ G\   (\op(\e_1,...,\e_n)) & = & \op(\mathcal{I}^\e\  G\  \e_1,..., \mathcal{I}^\e\  G\  \e_n)& \\
\mathcal{I}^\e\ G\  (\op(\e_1,...,\e_n)\ \At\ \x) & = & \x:=\op(\mathcal{I}^\e\  G\ \e_1,..., \mathcal{I}^\e\ G\ \e_n)& \\
\mathcal{I}^\e\ G\ (\lambda\p\!:\!\!\T.\e) & = &\lambda\pp.\  \mathcal{I}^\e\  \G\ \e\ \\
\mathcal{I}^\e\ G\ (\lambda\p\!:\!\!\T.\e\ \At\ \x) & = &\x :=\lambda\pp.\  \mathcal{I}^\e\  \G\ \e\ \\
\mathcal{I}^\e\ G\ (\f\ \e ) & = &  \f\ (\mathcal{I}^\e\ G\ \e)  \\
\mathcal{I}^\e\ G\ \La\e_1,...,\e_n\Ra & = & \La \mathcal{I}^\e\ G\ \e_1,...,\mathcal{I}^\e\ G\ \e_n\Ra \\
\mathcal{I}^\e\ G\ (\If\ \e\ \Then\  \e'\ \Else\ \e'' ) & = &\If\ \mathcal{I}^\e\ G\ \e\ \Then\  \mathcal{I}^\e\ G\ \e'\ \Else\ \mathcal{I}^\e\ G\ \e''  \\
\mathcal{I}^\e\ G\ (\Let\ \p \equiv \e\ \In\ \e') & = & \Let\ \pp\equiv\mathcal{I}^\e\ G\ \e\ \In\ \mathcal{I}^\e\ G\   \e'   \\
\end{array}
\\
$

\noindent
For the storable values we define $\mathcal{I}^\va\ G\  \va = \va$, if $\va$ is not a function value, and $\mathcal{I}^\va\  G\ (\lambda \p.\e)  =  (\lambda \pp.\ \mathcal{I}^\e\ G\ \e)$. Finally, if $\St=(\x_i=\va_i)_i$, then $\mathcal{I}\ G\ \St=(\x_i,\mathcal{I}\ G\ \ \va_i)_i$ and $\mathcal{I}\ G\ (\St,\e)=(\mathcal{I}\ G\ \St,\mathcal{I}\ G\ \e)$.

Some syntactic sugar is used inspired by the theoretical language Iswim \cite{reynolds}. $\e;\e'$ will be an abbreviation of $\Let\ \pp\equiv\e\ \In\ \e'$, where $FV\ \pp=\emptyset$.
Finally, if the function $\f=(\lambda \pp.\ \mathcal{I}^\e\ \e)$ satisfies $FV\ \pp=\emptyset$, then function calls of the form $\f\ (\mathcal{I}^\e\ \e) $ can be written $\mathcal{I}^\e\ \e;\ \f\ep$.

Below we show the imperative form $\mathcal{I}\ \!\{\vb,\z,\f,\Map\f,\Map,\ara,\vi,\n\}\ \!map$ of the program $map$, given in the section \ref{globalityandregions}. The presence in the initial store of variables with a non-significant value ($\arb,\z$ and others) is due to the need of the global program to force differentiated regions for particular fragments of the heap, so that the in-place update can be modeled through its collapse.

\medskip

\noindent\begin{tabular}{|c |}
\hline
$\begin{array}{ll}
\vb=0,\ \z=0,\ \f=0,\ \Map\f=0,\
\ara\ =\ \left\lbrace 0,1,2,3,4,5,6,7\right\rbrace,\ \vi=0,\ \n=0,\ \\
\Map\ = \lambda\ep.\ \Map\f:=(\lambda\La\Ra. \ \If\  \arb:=(\vi\!==\!\n) \\
\qquad\qquad\qquad\qquad\qquad\qquad\Then\ \La\ara,\vi,\n\Ra\\
\qquad\qquad\qquad\qquad\qquad\qquad\Else\ \Map\ \f;\ \!\La\ara[\vi]:=\Fun\ \! \z,\vi := (+1)\ \vi,\n\Ra);\ \!\Map\f\ \!\ep  \ \ \\\
\f := (\lambda \ep.\ \!\z := (+1)\ \z);\ \Map\ \f;\
\La\ara,\vi:=0, \n:=8\Ra;\ \Map\f\ \La\Ra\smallskip\\
\end{array}
$\\
\hline
\end{tabular}

\medskip

The example shown, which comes from the example we will study in section  \ref{globalityandregions}, forces globality on all variables.  For example, global typing allows forcing the globality of $\z$ by declaring $\f:\Pi\z.(\Int,\z)\!\!\rightarrow^{\{\z\}}\!\! \!(\Int,\z)$ and $\cdot[\ \cdot\ ] : ((\Array,\ara),(\Int,\vi))\rightarrow(\z,\Int)$. In this case (see definition of $\mathcal{I}$, case $\Let$) the evaluation does not perform the substitution, as can be seen in its transformation through the map $\mathcal{I}$ in the body of the function $map$. If in the text of the global program we were to write $\lambda \x\!\!:\!\!(\Int,\x).\ \!(+1)^\x\ \!\x$, and its type was $\Pi\x.(\x,\Int)\!\!\rightarrow^{\{\x\}}\!\!\! (\x,\Int)$, then its imperative form $\f := (\lambda \x.\ \!\x := (+1)\ \x)$ would force the substitution $\x\mapsto\z$ in the evaluation. That is, region abstraction in some sense also models passage by reference. Note that this option is not enabled for $\Let$-construct, which corresponds to the fact that we do not allow region abstraction in phrase $\Let$  of $\lambda[\Sigma^\sigma]$

It is necessary to clarify that the program $\Map$ that we will study in section  \ref{globalityandregions}, and that we use here to display the map $\mathcal{I}$, is an example completely at odds with linearity\footnote{Since it is still correct, it will allow us to observe its relationship with the regions program.}. It fully exploits the imperative potential, but its structure makes it clear that it is impossible to linearize most of its data. As we have already stated, our objective is to delve deeper into the relationship between globality and regions.

\section{Globality and regions}
\label{globalityandregions}

In $\lambda[\Sigma^\g]$, global variables, that is, those that occur in the initial store, generate a structure in the store that can be clearly replicated in a program with regions, and this occurs independently of the correctness of the global program\footnote{If the program is not correct with respect to the pure functional version, then the correlation between variables and regions is verified, but the relationship between the content is lost.}. Given a program $P$ in $\lambda[\Sigma^\g]$, which is globally well-typed in the type-context $\Gamma$, we can generate a program $\mathcal{R}^\Gamma\ \!\!P$ in $\lambda[\Sigma^\sigma]$ where there is a correlation between the global variables and the regions, and the value stored in the global variable at the end of the evaluation will turn out to be the last element stored in the region. The global context $\Gamma$ determines a structure of regions in the store, and the map $\mathcal{R}^\Gamma$ transforms global programs (expressions, values) into region programs (expressions, values).

More precisely, suppose that the program $P=(\St_0,\e)$ has substructural protection, and there exist $\Gamma,\T,\varphi$ (globals) satisfying $\Gamma\vdash\e\!:^\varphi\!\T$ and $\vdash \St_0\!:^\varphi\Gamma$. Assume that $\n^\Gamma\ \!\!\x$ is the region assigned to the global variable $\x\in dom\ \St_0$ (see sección \ref{elmapaR}). Let $\St_\R\ \!\vk$ denote the list $[\St\x_{\vk,1},...,\St\x_{\vk,n_\vk}]$, where $\R\ \!\vk=[\x_{\vk,1},...,\x_{\vk,n_\vk}]$. The following relation is verified: if $P\rightarrow^*(\St_1,\p)$ and $\mathcal{R}^\Gamma\ \!P\rightarrow^*(\R,\St,\p)$, then for every global variable $\x$ we have:

\medskip

$\St_1\ \x=last\ (\St_\R\ (\n^\Gamma\ \!\!\x))$

\medskip

Strictly speaking, this relationship occurs in basic types, since function-type values undergo modifications (in the evaluation of regions) because global variables are replaced by auxiliary variables of the corresponding regions. The example below illustrates this difference

We can interpret the regions program associated with a global program as generating a region for each global variable. Performing an in-place update in the global program corresponds to entering a new value into the region. In this way, the linearity property guarantees that the value of the global variable at the end of the evaluation is the last value entered into the region.

Obtaining the program with regions $\mathcal{R}^\Gamma\ \!\!P$ from the global $P$ is trivial (see sección \ref{elmapaR}), and basically consists of generating a place $\xi$ for each variable $\x$, and passing them as a region parameter in the functions.

This process is shown in the following figures\footnote{The examples were developed and tested in the Haskell prototype which can be found at https://github.com/hgramaglia/GloReg.}., along with the resulting store evaluation. We recall that in the examples we used the notation $\op^\g(\e_1,...,\e_n)$ as an abbreviation for $\op(\e_1,...,\e_n)\ \At\ \g$. We also write $\e_1\ \op^\g\ \e_2$ if the operation $\op$ is binary. We show the program on the left, and the store on the right with a representation that allows us to appreciate the structure defined by the regions.

For region evaluation, the data stored in the region is arranged from left to right (top to bottom for arrays or functions), with the rightmost (bottom) corresponding to the last entered. $\x_{\n,\vi}$ denotes the $i$-th element entered into the region. That is, $\R\n=[\x_{\n,1},...,\x_{\n,\vi},...]$. For example, we have $\x_{1,2}=\{2,3,3,4,5,6,7,8\}$.

In the expressión $\mathcal{R}^\Gamma\ map$ we take $\Gamma$ as:

\medskip

\noindent $\ara\!:\!(\Array,\ara), \vi\!:\!(\Int,\vi),\vk\!:\!(\Int,\vk),\vb\!:\!(\Int,\vb),\z\!:\!(\Int,\z), \f\!:\!(\Int,\f), \Map\f\!:\!(\Int,\Map\f)$.

\medskip

\noindent On the other hand, the value $\va_{7}$ is given by:

\medskip

\noindent $ \lambda \Lc\alpha,\iota,\kappa\Rc.\lambda \La\ara,\vi,\vk\Ra.\ \If\ (\vi==^{4}\vk) \ \Then\ \La\ara,\vi,\vk\Ra
\ \Else\smallskip $

$
\qquad\quad\Let\ \Map\f\equiv\Map\ [6]\ \x_{6,1}\ \In
\ \Map\ \Lc \alpha,\iota,\kappa\Rc\ \La\ara[\vi\rightarrow\x_{6,1}\ \Lc5\Rc\  (\ara[\vi]^{5})]^{\alpha},(+1)^{\iota}\ \vi,\vk\Ra$.

\medskip

\noindent
Note that the type of the global variables that occur only to the right of $\At$ (as well as their initial value in the store) is irrelevant, since they are only intended to reserve a region number in the initial allocation (e.g., $\arb$ is assigned region 4). These variables no longer occur in the regions program.

\medskip


\noindent\begin{tabular}{|c |c |}
\hline
$
\begin{array}{l}
\begin{array}{lcll}
\ara = \left\lbrace 1,...,8\right\rbrace,\ \vi=0,\ \n=0,\\
\vb=0,\ \z=0,\
\mathsf{f}\ =0,\ \Map\f\ =\ 0,\\
\Map\ = \ \lambda\f.\\
\ \ \lambda \La\ara,\vi,\vk\Ra:\La(\Array,\ara),(\Int,\vi),(\Int,\vk)\Ra.\ \\
\ \ \If\ (\vi==^\arb\vk) \ \Then\ \La\ara,\vi,\vk\Ra\\
\ \ \Else\ \Let\ \Map\f\equiv\Map\ \f\ \In\\
\quad\quad\ \ \Map\f\La\ara[\vi\rightarrow\Fun\ (\ara[\vi]^\z)]^\ara,(+1)^\vi\ \vi,\vk\Ra,\quad\\
\Let\ \f\equiv(\lambda \z:(\z,\Int) . (+1)^\z\ \z)^\f\ \In\\
\Let\ \Map\f \equiv \Map\ \mathsf{f}\ \In\ \Map\f\ \La\ara,0^\vi,8^\vk\Ra\\
\end{array}
\end{array}
$
&
$
\begin{array}{l}
\tiny\begin{array}{ll}
(\ara) &  \colv{\left\lbrace 2,3,4,5,6,7,8,9\right\rbrace}\qquad\quad \ \!\!\smallskip \\
(\vi)& \colv{8} \smallskip \\
(\vk) & \colv{8} \smallskip \\
(\arb) & \colv{\textsf{T}}\smallskip \\
(\z) & \colv{9}\smallskip \\
(\f)& \colv{\lambda\z.(+1)^\z\ \z}\smallskip\\
(\Map\f)& \colv{\va_{\Map\f}}
\\
\end{array}\\
\\
\tiny\begin{array}{l}
\va_{\Map\f}=_{def} \\
\lambda \La\ara,\vi,\vk\Ra.
\  \If\ (\vi==^\arb\vk) \ \Then\ \La\ara,\vi,\vk\Ra\\
\qquad \qquad\Else\ \Let\ \Map\f\equiv\Map\ \f\ \In\\
\quad \Map\f\La\ara[\vi\rightarrow\Fun\ (\ara[\vi]^\z)]^\ara,(+1)^\vi\ \vi,\vk\Ra,\quad\\
\end{array}
\end{array}
$\\
\hline
\end{tabular}\\

\medskip
\noindent\begin{tabular}{|c |c |}
\hline
$
\begin{array}{l}
\begin{array}{lcll}
\ara = \left\lbrace 1,...,8\right\rbrace,\
\vi=0,\ \vk=0,\\
\Map\ = \ \lambda\xi.\lambda\f.\\
\quad\lambda \Lc\alpha,\iota,\kappa\Rc.\lambda \La\ara,\vi,\vk\Ra\!\!:\!\!\La(\!\Array\!,\!\alpha\!),\!(\!\Int,\!\iota\!),\!(\!\Int,\!\kappa\!)\Ra\\
\qquad\If\ (\vi==^{4}\vk) \ \Then\ \La\ara,\vi,\vk\Ra\\
\qquad\Else\ \Let\ \Map\f\equiv\Map\ [\xi]\ \f\ \In\\
\qquad\Map\f\ \Lc \alpha,\iota,\kappa\Rc\\ \qquad\qquad\ \La\ara[\vi\rightarrow\Fun\ \Lc5\Rc\  (\ara[\vi]^{5})]^{\alpha},(+1)^{\iota}\ \vi,\vk\Ra,\\
\Let\ \f\equiv(\lambda \zeta. \z :(\Int,\zeta). (+1)^\zeta\ \z)^6\ \In\\
\Let\ \Map\f \equiv \Map\ [6]\ \mathsf{f}\ \In\\
\Map\f \ [1,2,3]\ \La\ara,0^2,8^3\Ra\\
\end{array}
\end{array}
$
&
$
\begin{array}{l}
\tiny\begin{array}{ll}
\\
(1) & \left\lbrace 2,2,3,4,5,6,7,8\right\rbrace,\\
& \left\lbrace 2,3,3,4,5,6,7,8\right\rbrace,\\
&\vdots\\
& \colv{\left\lbrace 2,3,4,5,6,7,8,9\right\rbrace}\smallskip \\
(2) &  0 \quad 1 \quad 2 \quad  3 \quad  4  \quad 5 \quad  6\quad  7 \quad  \colv{8} \smallskip \\
(3)& \colv{8} \smallskip \\
(4) & \textsf{F}\quad \textsf{F}\quad \textsf{F}\quad \textsf{F}\quad  \textsf{F} \quad \textsf{F}\quad  \textsf{F} \quad  \textsf{F}\quad  \colv{\textsf{T}}\smallskip \\
(5) & 1 \quad 2 \quad  2 \quad  3  \quad 3\quad  4\ \quad  4\quad  5 \\
& 5 \quad  6 \quad  6  \quad 7 \quad  7\quad   8\quad\ 8\quad \colv{8}\smallskip \\
(6)&\lambda\zeta\!.\!\lambda\z.(+1)^\zeta\ \!\z\ ...  \colv{\lambda\zeta\!.\!\lambda\z.(+1)^\zeta\ \!\z}\\
(7)& \va_7\ \mathsf{(defined\ bellow)}\\
&\vdots\\
&\colv{\va_7}

\end{array}
\end{array}
$\\
\hline
\end{tabular}\\

\medskip

We note that $map$, although correct, lacks substructural protection. To obtain this property, the  operator $\cdot[\ \cdot\ ]$ would have to be globalized (version $map_2$ of \cite{gramagliawlg}), in which case we would obtain the same memory management as seen in the version shown above. Another way to achieve substructural protection is to remove $\vb,\z$ from the set of global variables (version $map_1$ of \cite{gramagliawlg}). In this case, part of the evaluation would be handled in region 0 (pure part).

The region program obtained from the global version does not use the memory recovery mechanisms of the region language. Even when we incorporate these mechanisms, the memory deallocation regime of a global program (update in-place) is completely different from the memory release mechanism of a program with regions. A more efficient region program for $map$ is shown below.

\medskip

\noindent\begin{tabular}{|c |c |}
\hline
$
\begin{array}{l}
\begin{array}{lcll}
\ara = \left\lbrace 1,...,n\right\rbrace,\
\vi=0,\ \vk=0,\\
\Map= \lambda[\xi,\mu]\Ra.\lambda\f.\ \lambda \Lc\alpha,\iota,\kappa\Rc.\lambda \La\ara,\vi,\vk\Ra:\La(\Array,\alpha),(\Int,\iota),(\Int,\kappa)\Ra\\
\qquad\qquad\New\ \beta,\zeta.\ \If\ (\vi==^{\beta}\vk) \ \Then\ \La\ara,\vi,\vk\Ra\\
\qquad\qquad\qquad\qquad\Else\ \Let\ \Map\f\equiv\Map\ [\xi,\mu]\ \f\ \In\\
\qquad\qquad\qquad\qquad\qquad\Map\f\ \Lc \alpha,\iota,\kappa\Rc\  \La\ara[\vi\rightarrow\Fun\ \Lc\zeta\Rc\  (\ara[\vi]^{\zeta})]^{\alpha},(+1)^{\iota}\ \vi,\vk\Ra,\ \ \qquad\\
\New\ \xi,\mu.\  \Let\ \f\equiv(\lambda \zeta. \z :\zeta\ \Int. (+1)^\zeta\ \z)^\xi\ \In\\
\qquad\qquad\Let\ \Map\f \equiv \Map\ [\xi,\mu]\ \mathsf{f}\ \In\ \Map\f \ [1,2,3]\ \La\ara,0^2,8^3\Ra\\
\end{array}
\end{array}
$
\\
\hline
\end{tabular}\\

\medskip

The context  $\Gamma $ that types the program satisfies: $\Gamma=\ara:(\Array,1),\ \vi:(\Int,2),\ \n:(\Int,3),\ \Map:(\T_\Map,4)$, where:

\smallskip

\noindent $\T_\Map=\Pi[\xi,\mu].(\Pi.\zeta.(\Int,\zeta)\rightarrow^{\{\zeta\}}(\Int,\zeta) ,\xi)\rightarrow^{\{\mu\}}$

\smallskip

$(\Pi[\mu,\alpha,\iota].\La(\Array,\alpha),(\Int,\iota),(\Int,\kappa)\Ra\rightarrow^{\{\mu,\alpha,\iota,\kappa\}}\La(\Array,\alpha),(\Int,\iota),(\Int,\kappa)\Ra,\mu)$

\medskip

Finally, we show a version of $map$ in which the presence of pure algorithm components (qualifier $\lo$) generates a part of the heap in the regions program that is not related to any global variable (region 0). This data will not be eliminated by the substructural version or the global version.

In the next figure, $\va_\g$ denotes the value $\lambda\La\ara,\vi,\n\Ra.\If\ \vi==\n\ \Then \
\La\ara,\vi,\n\Ra\ \Else$ $\Let\ \Map\f\equiv\Map\ \x_1\ \In\ \Map\f\La\ara[\vi\rightarrow\x_1\ (\ara[\vi])]^\ara,(+1)^\vi\ \vi,\vk\Ra$ (remember that $\vi==\n$ is a short notation for $\vi==^\lo\n$).

For both the heap in the global evaluation and region 0 in the evaluation of regions, we will explicitly show the auxiliary variables, using a numbering to indicate the order in which they are introduced. We recall that $\e\ \op\ \e'$ (resp. $(\lambda\p.\e)$) is a short notation for $\e\ \op^\lo\ \e'$ (resp. $(\lambda\p.\e)^\lo$).

It should be noted in this global version the existence of a precarious globality in the variable $\z$, which is limited to the scope of $\lambda \z\!\!:\!\!(\Int,\lo) . (+1)^\z\ \z$. The qualifier $\z$ in $(+1)^\z\ \z$ produces the rewriting of the variables $\x_5,\x_9,...,\x_{33}$, which establishes an appreciable difference between the heap of the global evaluation, and region 0 of the evaluation by regions. That is, globalization produces the ``collapse'' of small subregions of region 0

\medskip


\noindent\begin{tabular}{|c |c |}
\hline
$
\begin{array}{l}
\begin{array}{lcll}
\ara = \left\lbrace 1,...,8\right\rbrace,\ \vi=0,\ \n=0,\\
\Map= \lambda\f.\lambda \La\ara,\vi,\vk\Ra\!:\!\La(\Array,\ara),(\Int,\vi),(\Int,\vk)\Ra.\!\!\! \\
\ \ \If\ (\vi==\vk) \ \Then\ \La\ara,\vi,\vk\Ra\\
\ \ \Else\ \Let\ \Map\f\equiv\Map\ \f\ \In\\
\quad\quad\ \ \Map\f\La\ara[\vi\rightarrow\Fun\ (\ara[\vi])]^\ara,(+1)^\vi\ \vi,\vk\Ra,\\
\Let\ \f\equiv(\lambda \z:\lo\ \Int . (+1)^\z\ \z)\ \In\\
\Let\ \Map\f \equiv \Map\ \mathsf{f}\ \In\ \Map\f\ \La\ara,0^\vi,8^\vk\Ra\\
\end{array}
\end{array}
$
&
$
\begin{array}{l}
\tiny\begin{array}{ll}
\\
\mathsf{HEAP} \\
\x_1=\lambda\z.(+1)^\z\ \z,\ \x_2= \va_{\m},\\
\x_3=\mathsf{F},\ \x_4= \va_{\g},\ \
\col{\x_5 = 2},\\
\x_7=\mathsf{F},\ \x_8=\va_{\g},\ \ \col{\x_9 = 3},\\
\x_{11}=\mathsf{F},\ \x_{12}=\va_{\g},\ \x_{13} = 4,\\
...\\
\x_{31} =\mathsf{F},\ \x_{32 }=\va_{\g},\ \col{\x_{33} = 9},\\
\x_{35} =\mathsf{T},\smallskip\\
\\
(\ara) \quad  \colv{\left\lbrace 2,3,4,5,6,7,8,9\right\rbrace}\qquad\quad\ \ \smallskip \\
(\vi)\quad \colv{8} \smallskip \\
(\vk)\quad \colv{8} \smallskip \\
\\
\end{array}
\\
\end{array}
$\\
\hline
\end{tabular}\\

\medskip

In the next figure, which represents the region program associated with the previous program, $\va_\vr$ denotes the value:
\medskip

\noindent $ \lambda \Lc\alpha,\iota,\kappa\Rc.\lambda \La\ara,\vi,\vk\Ra.\ \If\ (\vi==^{0}\vk) \ \Then\ \La\ara,\vi,\vk\Ra
\ \Else\smallskip $

$
\qquad\quad\Let\ \Map\f\equiv\Map\ [0]\ \x_{0,1}\ \In
\ \Map\ \Lc \alpha,\iota,\kappa\Rc\ \La\ara[\vi\rightarrow\x_{0,1}\ \Lc0\Rc\  (\ara[\vi]^{0})]^{\alpha},(+1)^{\iota}\ \vi,\vk\Ra$.

\medskip

\medskip

\noindent\begin{tabular}{|c |c |}
\hline
$
\begin{array}{l}
\begin{array}{lcll}
\ara = \left\lbrace 1,...,8\right\rbrace,\
\vi=0,\ \vk=0,\\
\Map = \lambda\ep.\lambda\f.\\
\ (\lambda \Lc\alpha,\iota,\kappa\Rc.\lambda \La\ara,\vi,\vk\Ra\!:\!\La(\Array,\alpha),(\Int,\iota),(\Int,\kappa)\Ra\!\!\!\\
\qquad\If\ (\vi==^{0}\vk) \ \Then\ \La\ara,\vi,\vk\Ra\\
\qquad\Else\ \Let\ \Map\f\equiv\Map\ [0]\ \f\ \In\\
\qquad\Map\ \Lc \alpha,\iota,\kappa\Rc\\ \qquad\qquad\ \La\ara[\vi\rightarrow\Fun\ \Lc0\Rc\  (\ara[\vi]^{0})]^{\alpha},(+1)^{\iota}\ \vi,\vk\Ra)^0,\\
\Let\ \f\equiv(\lambda \ep. \z :(\Int,0). (+1)^0\ \z)^0\ \In\\
\Let\ \Map\f \equiv \Map\ [0]\ \mathsf{f}\\
\In\ \Map\f \ [1,2,3]\ \La\ara,0^2,8^3\Ra\\
\end{array}
\end{array}
$
&
$
\begin{array}{l}
\tiny\begin{array}{ll}
(0)\ (\textsf{Here } \x_i=\x_{0,i})\\
\qquad\x_{1}\!=\!\lambda\ep\lambda\z.(+1)^0 \z,\ \x_2\!=\! \va_{\m},\\
\qquad\x_3=\mathsf{F},\ \x_4= \va_{\vr},\ \
\col{\x_5 = 1},\\
\qquad\col{\x_6 = 2},\ \x_7=\mathsf{F},\ \x_8=\va_{\vr},\\
\qquad \col{\x_9 = 2,\ \x_{10} = 3},\
\x_{11}\!=\!\mathsf{F},\\
\qquad\x_{12}=\va_{\vr},\ \x_{13} = 3,\\
\qquad\x_{14}=4 ,\\
\qquad\ ...\\
\qquad\x_{31} =\mathsf{F},\ \x_{32 }=\va_{\vr},\ \col{\x_{33}\! =\! 8},\\
\qquad\col{\x_{34} = 9},\
\x_{35} =\mathsf{T},\smallskip\\
\\
(1) \ \left\lbrace 2,2,3,4,5,6,7,8\right\rbrace,\\
\qquad \left\lbrace 2,3,3,4,5,6,7,8\right\rbrace,\\
\qquad...\\
\qquad \colv{\left\lbrace 2,3,4,5,6,7,8,9\right\rbrace}\smallskip \\
(2) \ \  0 \ \ 1 \ \ 2 \ \  3 \quad  4  \ \  5 \ \   6\ \  7 \ \  \colv{8} \smallskip \\
(3)\ \  \colv{8} \smallskip \\
\end{array}
\end{array}
$\\
\hline
\end{tabular}\\

\medskip

\section{The map $\mathcal{R}$}
\label{elmapaR}

In this section we will give the formal definition of the map $\mathcal{R}^\Gamma\ (\St,\e)$ introduced in section \ref{globalityandregions}. It will be defined for globally well-typed $\lambda[\Sigma^\g]$-programs $(\St,\e)$. That is, there exist global $\Gamma,\T$ such that $\vdash\St:\Gamma$ and $\Gamma\vdash\e:\T$.

Before defining the map $\mathcal{R}$ we must introduce some notation. We will assign the initial store $\Gamma$ a distribution in regions as follows. Global variables will have a region number according to their order of appearance, and local variables will be assigned to region 0. This is reflected in the following definition. If $\x$ is a variable declared in the context $\Gamma$, then we denote by $\n^\Gamma\ \!\!\x$ the integer defined as follows. If $\Gamma\x=(\B,\x)$ then $\n^\Gamma\ \!\!\x$ is determined by the following condition: the $(\n^\Gamma\ \!\!\x)$-th global variable of basic type declared in $\Gamma$ is $\x$. Otherwise we define $\n^\Gamma\ \!\!\x=0$.

The association of global variables with regions is done through the map $q^\g_s$, defined as follows. Here $s$ is a set of variables.\footnote{Initially $s$ will represent the variables that in the current environment have a specific (non-abstract) region assigned.}. We assume a map $\x\rightarrow\alpha_\x$ that uniquely associates variables with region symbols.

\medskip
$\begin{array}{rllll}
q^{\lo}_s &  = & 0 & \smallskip\\
q^{\x}_s &  = & \alpha_\x & (\x\notin s)\smallskip\\
q^{\x}_s &  = & \n^\Gamma\x & (\x\in s)\\
\end{array}
$

\medskip

If  $\p$ is a pattern, and $s$ is a set of variables, then we denote by $\q^\T_{s}\ \!\!\p$ the pattern of places defined as follows.

\medskip
$\begin{array}{rllll}
\q^{(\Pt,\g)}_s\  \!\x &  = & q^{\g}\  \x& (\Pt=\B,\Pi\p.\T\rightarrow^\varphi\T')\smallskip\\
\q^{\ep}_s\  \!\x &  = & 0& \smallskip\\
\q^{\La\T_1,...,\T_n\Ra}_s\  \!\La\p_1,...,\p_n\Ra &  = & \q^{\T_1}_s\  \!\p_1,...,\q^{\T_n}_s\  \!\p_n&(n\neq 0) \\
\end{array}
$

\medskip

In particular, we denote by  $\alpha_\p^{\T}$ the subsequence of effect variables (i.e. variables of the form  $\alpha_{\_})$ of  $\q^\T_{[]}\p$. We define below the map $\mathcal{R}$ for the expressions. The map $\p^\Gamma$ is defined in section \ref{sistemadetipoglobal}.

\medskip
\noindent$
\begin{array}{rclll}
\mathcal{R}^\Gamma_s\  \x & = &\x &\smallskip\\
\mathcal{R}^\Gamma_s\  (\op(\e_1,...,\e_n)\ \At\ \g) & = & \op(\mathcal{R}^\Gamma_s\ \e_1,..., \mathcal{R}^\Gamma_s\ \e_n)\ \At\ q^{\g}_s&\smallskip\\
\mathcal{R}^\Gamma_s\ \xs\ (\If\ \e\ \Then\  \e'\ \Else\ \e'' ) & = &\If\ \mathcal{R}^\Gamma_s\  \e\ \Then\  \mathcal{R}^\Gamma_s\ \e'\ \Else\ \mathcal{R}^\Gamma_s\ \e'' \smallskip \\
\mathcal{R}^\Gamma_s\  \La\e_1,...,\e_n\Ra & = & \La \mathcal{R}^\Gamma_s\ \e_1,...,\mathcal{R}^\Gamma_s\ \e_n\Ra\smallskip \\
\mathcal{R}^\Gamma_s\ (\f\ \e ) & = &  \f\ [\q^\T_s\ \!\! (\p^\Gamma\e)]\ (\mathcal{R}^\Gamma_s\ \e)\quad\qquad(\Gamma\f=(\Pi\p.\!\T\!\!\rightarrow^\varphi\!\T'\!\!,\!\g)) \smallskip \\
\mathcal{R}^\Gamma_s\ (\Let\ \p \equiv \e\ \In\ \e') & = & \Let\ \p\equiv\mathcal{R}^\Gamma_s\  \e\ \In\ \mathcal{R}^{\Gamma,[\p:\T]}_{s}\ \e'\qquad\qquad\ \ (\Gamma\vdash\e\!:^\varphi\T) \smallskip  \\
\end{array}
$

\medskip

\noindent Finally, for abstraction we have the equation:

\medskip
$
\begin{array}{rclll}
\mathcal{R}^\Gamma_s\  (\lambda\p\!:\!\!\T.\e\ \At\ \g) & = & \lambda \alpha_\p^\T.\lambda\p\!:\!\mathcal{R}^{\Gamma,[\p:\T]}_{s'} \T.\mathcal{R}^{\Gamma,[\p:\T]}_{s'} \e\  \At\ q^\g_s
\end{array}
$

\medskip
\noindent where $s'=(s-\{\x\in FV\ \!\p:\n^\Gamma\x\neq0 \})\cup\{\x\in FV\ \!\p:\lo([\p\!:\!\!\T]\x)\}$.

\medskip

\noindent
For storable values we define
$\mathcal{R}^\Gamma_s\  \!(\x,\va) = (\x,\va)$, if $\va$ is not a function value, and

\medskip

$\mathcal{R}^\Gamma_s\ (\x,\lambda \p.\e)  =  ( \x, \lambda \alpha_\p^\T.\lambda\p.\mathcal{R}^{\Gamma,[\p:\T]}_{s'} \e)$,

\medskip

\noindent where $\Gamma\x=(\Pi\p.\!\T\!\rightarrow^\varphi\!\T'\!,\g)$, and $s'$ is defined above. If $\St=(\x_i=\va_i)_i$, then

\medskip

$\mathcal{R}^\Gamma\ (\St,\e)=((\x_i,\mathcal{R}^\Gamma_s\ \! \va_i)_i,\mathcal{R}^\Gamma_s\ \!\e)$,

\medskip

\noindent where $s=\{\x_i \}_i$.
Finally, the region context $R^\Gamma$ is defined by:

\medskip
$
\begin{array}{rllll}
R^\Gamma &  = & \sum [(\n^\Gamma\x,\x)\ |\ \x\in\Gamma].\\
\end{array}
$

\medskip

\section{Conclusions: linearity, globality and regions}
\label{conclusiones}

One of the current challenges in programming language research is providing tools for reasoning about the behavior of their codes in the presence of imperative operations such as in-place updates and memory deallocation. Type systems based on Girard's linear logic \cite{girard,lafont,wadler} and the discipline of types and effects developed by Gifford and Lucassen \cite{gifford}, and refined by Jouvelot, Talpin, and Toften \cite{jouvelot,jouvelottofte,tofte}, emerged as powerful tools to address this challenge.

The global language $\lambda[\Sigma^\g]$, in addition to providing a conceptually clear way to incorporate in-place update into a functional language, provides a simple conceptualization for global variables. This language has the virtue of positioning itself between languages that abstractly model imperative operations and the imperative language itself. In this article we have modified the language presented in \cite{gramagliawlg} by making global variables explicit as effects, in order to make their relationship with the region language more evident. In the region system, effects are necessary to ensure safety when deleting memory blocks due to the presence of function-type values. In the original language $L^1[\Sigma^\g]$ presented in \cite{gramagliawlg}, functions are not subject to the special memory management regime that each system proposes.

In \cite{gramagliawlg} we saw how the languages $L^1[\Sigma^\q]$ and  $L^1[\Sigma^\g]$ allow, by relating $\Sigma^\q$ and $\Sigma^\g$, a simple definition of the concept  substructural protection: weak-linear typing guarantees the correctness of the imperative program.
On the other hand, we saw in this article how the update-in-place can be modeled by the structure of regions of the  heap, determined by the global variables and the idea of collapsed regions: when a new value enters the region, the existing one is destroyed. In this work we extend the special memory management modalities (global and regions) to functions, establishing a more general framework than \cite{gramagliawlg}.
This preliminary work has as its starting point the strategy of concentrating the special attributes of a type system  (substructurality and globality) in the signature. In particular, tuples are removed from the set of storable values, which forces the use of patterns in the lambda abstraction. Although this introduces some awkward notational handling, it is essential for simplifying the bureaucracy of syntax in the substructural language. We believe that this strategy, although it results in a decrease in the expressive power of the original languages, enhances the tools by granting them conceptual clarity and economy of syntax. In \cite{gramagliawlg}, this strategy also favors the automatic obtaining of substructural versions of a program.

Limiting ourselves to the languages $L^1[\Sigma^\q]$ (linear), $L^1[\Sigma^\g]$ (global) and  $L^1[\Sigma^\sigma]$\footnote{That is, excluding the functions of the region regime.} (regions) allows us to observe what each system contributes to the understanding the problem of incorporating imperative elements into a functional language. Even so, these languages constitute a strange puzzle with missing pieces, that has not yet been able to constitute a completely adequate model to address the aforementioned problem.

\end{document}